\documentclass[runningheads,a4paper]{llncs}
\pdfoutput=1
\usepackage{tikz}
\usetikzlibrary{shapes,arrows,shadows,spy,shapes.geometric,positioning,decorations.pathreplacing}
\usepackage{pgf}
\usepackage{graphicx}
\usepackage{xcolor,colortbl}
\usepackage{multirow}
\usepackage{changepage}
\usepackage{lmodern}
\usepackage{amsmath,bm,times} 
\usepackage{verbatim}
\usepackage[english]{babel}
\usepackage{lscape}
\usepackage{subfig}
\usepackage{booktabs}
\usepackage{paralist}
\usepackage{csquotes}
\usepackage[T1]{fontenc}
\usepackage{microtype}
\usepackage[math]{blindtext}

\usepackage[%
rm={oldstyle=false,proportional=true},%
sf={oldstyle=false,proportional=true},%
tt={oldstyle=false,proportional=true,variable=true},%
qt=false%
]{cfr-lm}

\usepackage[
bookmarks=false,
breaklinks=true,
colorlinks=true,
linkcolor=black,
citecolor=black,
urlcolor=black,
pdfpagelayout=SinglePage
]{hyperref}
\usepackage[all]{hypcap}

\usepackage[capitalise]{cleveref}
\crefname{section}{Sect.}{Sect.}
\Crefname{section}{Section}{Sections}
\crefname{figure}{Fig.}{Fig.}
\Crefname{figure}{Figure}{Figures}

\usepackage{xspace}

\DeclareFontFamily{U}{MnSymbolC}{}
\DeclareSymbolFont{MnSyC}{U}{MnSymbolC}{m}{n}
\DeclareFontShape{U}{MnSymbolC}{m}{n}{
    <-6>  MnSymbolC5
   <6-7>  MnSymbolC6
   <7-8>  MnSymbolC7
   <8-9>  MnSymbolC8
   <9-10> MnSymbolC9
  <10-12> MnSymbolC10
  <12->   MnSymbolC12%
}{}
\DeclareMathSymbol{\powerset}{\mathord}{MnSyC}{180}

\makeatletter
\g@addto@macro{\UrlBreaks}{\UrlOrds}
\makeatother

\tikzstyle{block} = [draw, fill=white!20, rectangle,align=center,minimum height=4em, minimum width=6em, drop shadow,rounded corners]
 
\tikzstyle{texblock} = [draw,draw=none, rectangle,align=center,minimum height=4em, minimum width=6em]

\tikzstyle{dist} = [draw, fill=gray!20, circle, node distance=1cm]
\tikzstyle{input} = [coordinate]
\tikzstyle{output} = [coordinate]
\tikzstyle{pinstyle} = [pin edge={to-,thick,black}]

\usepackage{url}
\urldef{\mailsa}\path|{oe26, nw91}@le.ac.uk|
\newcommand{\keywords}[1]{\par\addvspace\baselineskip
\noindent\keywordname\enspace\ignorespaces#1}

\begin{document}


\title{Finding  Clustering Configurations to Accurately Infer Packet Structures from Network Data}

\author{Othman Esoul \& Neil Walkinshaw}
\institute{Department of Computer Science\\
          University of Leicester\\
          LE1 7RH, Leicester, UK. \\
         \mailsa}

%

\maketitle

\begin{abstract}
Clustering is often used for reverse engineering network protocols from captured network traces. The performance of clustering techniques is often contingent upon the selection of various parameters, which can have a severe impact on clustering quality. In this paper we experimentally investigate the effect of four different parameters with respect to network traces. We also determining the optimal parameter configuration with respect to traces from four different network protocols.
Our results indicate that the choice of distance measure and the length of the message has the most substantial impact on cluster accuracy. Depending on the type of protocol, the $n$-gram length can also have a substantial impact.  
\end{abstract}


\section{Introduction}\label{sec:intro}

Protocol reverse-engineering (or protocol inference) is concerned with the challenge of inferring a specification of a network protocol specification from traces of network data. Inferred protocols can be valuable in a multitude of scenarios, especially in the contexts of security and testing. Inferred protocols can be used to derive novel test cases for black-box fuzzing \cite{walkinshaw2010increasing,aarts2014improving}, can be used to interact with and explore botnets \cite{caballero-2009}, or can be built into intrusion detection / supervisor frameworks \cite{comparetti-2009}.

A crucial step for any inference technique is to infer the packet structures from the data, so that it is possible to interpret a data stream as a sequence of packets. Most current approaches \cite{marshall-2004,pext-2007,discoverer-2007,comparetti-2009,Wang-2012} identify common patterns within the data by way of an unsupervised Data Mining technique known as \emph{clustering} \cite{Halkidi-2001}. Clustering can empirically elucidate the "natural", unknown and ideally interesting groups of messages within the captured network trace. These groups can then be used to identify the possible structures of message types implemented in the protocol. 

Most network protocol inference techniques that involve clustering follow a common sequence of steps, but vary substantially in terms of the specific methods or parameters that they adopt with respect to clustering. For example, they might pre-process the data in different ways (e.g. limit messages to the first, 32 or 64 bytes, or fragment the message as $n$-grams). They might adopt different combinations of ``distance measures''. They might be tailored towards text-based protocols or binary ones. 

Most of the empirical results are presented with respect to a fixed configuration of clustering parameters. However, the sensitivity of clustering algorithms to their parameters  \cite{Anderberg-1973} suggests that performance could vary significantly, depending on factors such as the type of protocol, the choice of distance measure, the amount of data, etc. Accordingly, this paper explores the following questions:
\begin{itemize}
\item RQ1  What is the effect of each variable on clustering accuracy?
\item RQ2  What is the optimal configurations for clustering?
\end{itemize}

To answer the questions we have carried out an empirical study. This assesses the impact of four different parameters with respect to four real-world protocols. The chosen variables in the experimental study are: the \emph{length of the message}, \emph{size of the sample}, \emph{length of the $n$-gram} (a message tokenisation approach used extensively by several applications), and the choice of \emph{distance measure} (often required by clustering algorithms). The network protocols included in this study are: the \emph{Trivial File Transfer Protocol} (TFTP), \emph{Domain Name Service} (DNS), \emph{Server Message Block} (SMB), and \emph{Hyper-Text Transfer Protocol} (HTTP).

In this  study, we have quantified the effect of each variable on clustering accuracy and have used this to identify an optimal configuration for clustering. Our results show that the choice of the distance measure have the \emph{largest} effect on clustering accuracy, followed by the length of the message. Our results also indicate that combining the \emph{Ball-Hall} internal clustering validation index with the \emph{Braun-Blanquet} distance measure achieves results that are consistently better than other combinations.

\section{Background and Motivation} \label{sec:backgr}
In this section we begin with a general overview of protocol reverse engineering techniques. We then present the general sequence of steps that most approaches tend to adopt for clustering packet data. The section concludes with a discussion of the motivation of our work.

Network protocol specifications are the backbone of several security applications \cite{bro-1999,fuzzing-book-2007,Caballero07fig,unexpected-2006,Kim-2008,binpac-2006}. Given an undocumented protocol (e.g., SMB, Skype), the goal of protocol reverse engineering is to extract the \emph{message format}, which captures the structure of all messages that comprise the protocol, and the protocol \emph{state machine}, which captures the valid sessions (message sequences) of the protocol. There are two common approaches for inferring protocol specifications: (1) by \emph{reverse engineering} protocol implementations (e.g., sever-side analysis of executables while processing messages), or (2) by \emph{analysing network traffic}. In this paper, we focus on the latter approach.

\begin{figure}
\begin{center}
\scalebox{0.7}
{
\begin{tikzpicture}[auto]
    \node [block, pin={[pinstyle] below:Network Traces}, node distance=2.3cm, text width=2.0cm] (TC) {Traffic Classification};
    \node [block, right of=TC] [node distance=3.8cm, text width=2.0cm] (MP) {Message Preprocessing};
    \node [block, right of=MP] [node distance=3.8cm, text width=2.0cm] (MC) {Message Clustering};
    \node [block, right of=MC][node distance=3.8cm, text width=2.0cm] (MS) {Message Alignment};
	\node [output, below of=MS] [node distance=1.5cm] (EX2) {}; 

	\draw [->] (TC) -- node[name=kln1,text width=2.5cm, align=center] {Messages} (MP); 
	\draw [->] (MP) -- node[name=kln2] {Features} (MC);     
    \draw [->] (MC) -- node[name=kln3] {Clusters} (MS);
    \draw [->] (MS) -- node[name=kln3] {Msg. Formats} (EX2);
\end{tikzpicture}
}
\end{center}
\caption{Common sequence of steps for network-based protocol reverse engineering.}
\label{fig_A}
\end{figure}
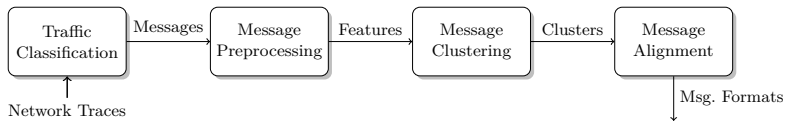

\subsubsection{Common Approach.}
Figure \ref{fig_A} provides a high-level flow-chart of the the common sequence of steps that tend to be adopted by most traffic-based reverse engineering techniques to infer the message structure. Typically, the approach consists of following steps: \emph{traffic classification}, \emph{message preprocessing}, \emph{message clustering}, and \emph{message alignment}. 

In the \emph{traffic classification} step only messages that belong to target protocol are extracted for analysis. There are several ways to accomplish this task \cite{Kim-2008}. 

The \emph{message preprocessing} step prepares protocol messages for clustering. This tends to involve data cleansing (e.g. filtering out irrelevant data) and dimensionality reduction \cite{Maimon_2005} (reducing the number of features in terms of which the messages are to be clustered). 

Typically, application protocols involve multiple different types of messages where each type has it own format. The \emph{clustering} step serves to identify the possible types of these messages. This is achieved by partitioning the protocol messages into multiple distinct groups where messages in one cluster are of the same type following the same format. 

Finally, The last step in the process is normally the message alignment step. \emph{Sequence alignment} algorithms are often used (e.g., Needleman Wunsch algorithm  \cite{needleman1970}) to align protocol messages of the same type. The sequence alignment algorithm takes as input two similar protocol messages and align them, exposing the structural aspects of field similarities, differences, and gaps (if both messages have different lengths).

\subsubsection{Motivation.}
The common approach discussed above consists of multiple  steps. Crucially, the choices that are made with respect to choosing the parameters for each of these steps can have a significant impact on the accuracy of the resulting inference results. The ideal choices may depend to an extent upon the characteristics of the network data (the amount of data available, the nature of the data (e.g. whether it is a text or binary protocol). Moreover, these factors are not independent; the effect of choosing a particular approach to tokenising the network data may be dependent on the choice of distance measure used to cluster the data, and might also depend on the amount and nature of the network data.

Choosing a suitable clustering configuration is ultimately a complex process. However, there is a dearth of guidance that can indicate how to choose different settings. Most protocol inference approaches are evaluated with respect to a static configuration. This is what motivates the work presented in this paper: to provide an experimental framework, along with some empirical data that can be used to guide the choice of suitable clustering configurations for packet extraction.

\begin{figure}
\begin{center}
\scalebox{0.6}
{
\begin{tikzpicture}[auto]
    \node [block, pin={[pinstyle]below:Network Traces}, node distance=2.3cm, text width=2.0cm] (TC) {Message Classification};
    \node [block, right of=TC] [node distance=3.0cm, text width=2.0cm] (MP) {Message Manipulation};
    \node [block, right of=MP] [node distance=3.0cm, text width=2.0cm] (MT) {Message Tokenisation};
    \node [block, right of=MT] [node distance=3.0cm, text width=2.0cm] (FS) {Feature Selection};
    \node [block, right of=FS] [node distance=3.0cm, text width=2.0cm] (MC) {Message Clustering};
    \node [block, right of=MC] [node distance=3.0cm, text width=2.0cm] (CV) {Clustering Validation};
    \node [output, below of=CV] [node distance=1.3cm] (EX2) {}; 

	\draw [->, double] (TC) -- node[name=kln1,text width=2.5cm, align=center] {} (MP); 
	\draw [->, double] (MP) -- node[name=kln2] {} (MT);     
    \draw [->, double] (MT) -- node[name=kln3] {} (FS);
    \draw [->, double] (FS) -- node[name=kln3] {} (MC);
    \draw [->, double] (MC) -- node[name=kln3] {} (CV);
    \draw [->, >=latex] (CV) -- node[name=kln3] {Validation Scores} (EX2);
\end{tikzpicture}
}
\end{center}
\caption{Experimental Framework for message clustering and validation}
\label{fig_B}
\end{figure}
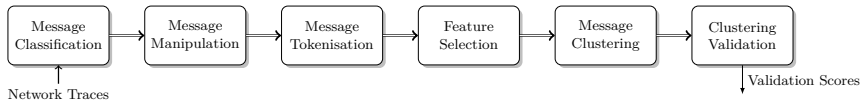

Therefore, it would be helpful if we could quantitatively assess how \emph{large} or \emph{small} the effect of those variables on clustering accuracy, and whether we could predict the best combination of these variables that enable us to achieve the best possible clustering.
\section{A Modular Message Clustering Framework}
In this section, we present a framework that enables us to provide answers to the above questions. The framework provides an intuitive, extensible basis for improving clustering and protocol inferencing in general. It takes the common stages outlined in Figure \ref{fig_A} and use to provide a controllable modular environment for clustering. This can be easily used to generate different message clustering configurations as compositions of various stages and parameters. Because we are especially interested in clustering (and steps lead up to clustering), we have integrated \emph{clustering validation} step. This is to be able to evaluate clustering results and guide the inference process. The framework takes as input captured network messages and produces clustering validation scores. 

The framework is shown in Figure \ref{fig_B} and explained in more detail below. Whenever a stage subject to parameter choices, these are listed in bold.

\subsubsection{Traffic Classification.}
The traffic classification method used in this step is the port-based method \cite{Kim-2008}. Port-based traffic classification  relies upon the use of port numbers in the \emph{transport layer} to filter network traffic. Typically, each protocol has standard port number(s) to represent that application of the protocol. We assume that the collected traffic is healthy (no malformed packets) and there is no misuse of port numbers, e.g. use of non-standard port numbers for communication \cite{unexpected-2006}.

\subsubsection{Message Preprocessing.}
We have divide the message preprocessing stage into the following steps:

\paragraph{Sample Manipulation.} This step accomplishes two tasks: First, to extract only data that belongs to the application layer protocol, i.e. data that belongs to the \emph{transport layer}, \emph{network layer} and \emph{link layer} are discarded. Second, this step is also utilised to assign (manipulate) different sample sizes and message lengths according to different sizes and lengths. Parameters of this step: \textbf{Sample Size} \& \textbf{Message Length}. 

\paragraph{Message Tokenisation.} \label{eq-ng} We use $n$-grams \cite{Cavnar94,Wang-2012} to tokenise protocol messages. An $n$-gram is a subsequence of $n$ consecutive characters from a longer sequence. The $n$-gram's approach does not require protocol field delimiters to be predefined to the tokeniser. Normally, the result of this step is a large number of $n$-grams. The number of $n$-grams which can be generated from a message of length $m$ using an $n$-gram of length $n$ can be calculated from the following equation: $m-n+1$ where $(n \leq m)$. Parameters of this step: \textbf{$N$-gram Length}. 

\paragraph{Feature Selection.} Messages from the same type normally have similar \emph{$n$-gram frequency distributions} \cite{Wang-2012}, therefore, we use the $n$-gram occurrences as a feature to distinguish between protocol messages (frequencies of the $n$-grams are counted in relation to their messages). To normalise the amount of contribution of each $n$-gram, we apply the \emph{Term Frequency-Inverse Term Frequency} (TF/IDF) as a weighting scheme \cite{Rieck-2008}. Also, we eliminate $n$-grams that carry no discriminative features. Since the generated feature space is mostly sparse, we remove $n$-grams which occur very infrequently (i.e. sparse $n$-grams); retaining only the common $n$-grams. We set the maximum sparseness allowed for $n$-grams to be retained to a certain percentage for the entire experiment. 

\subsubsection{Message Clustering.}
In this step we cluster similar protocol messages into distinct clusters. We use an \emph{agglomerative hierarchical clustering} algorithm with \emph{complete linkage} clustering criteria \cite{Jain-1988}. Separate clusters are obtained by cutting the generated tree (dendrogram) at a given height. Throughout the experiment, we fix the cutting height to a certain level. Agglomerative hierarchical clustering requires a distance measure, the selected distance measures used for the experiment are explained in the evaluation section. Parameters of this step: \textbf{Distance Measure}.

\subsubsection{Clustering Validation.}
Clustering validation is the process of evaluating the result of a clustering algorithm. In general, cluster validation can be divided into two categories, \emph{external  validation} and \emph{internal  validation}. External validation measures require the actual "true" classes to be known a-priori. Internal measures evaluate the goodness of clustering based on internal geometrical aspects of the data (e.g., compactness and separation) without any external information. 

We validate clustering results using external and internal clustering validation measures of choice. Through the external validation measure, clustering is validated by comparing the produced partitions from the clustering algorithm with the \emph{ground truth} partitions. Instead of manually extracting message types from the formal documentations of the protocols, we use of-the-shelf network analyser that is capable of correctly parse the network traffic of the protocol to identify and label message types. We use \emph{tshark} network analyser \cite{tshark-2016} (a command-line version of Wireshark) to automatically identify and extract true message labels to be provided to the external measure.

\section{Evaluation}
This section, consists of two parts. First, we present the experimental set-up that describes the experimental subjects and variables to be part of the experiment. The second part presents the methodology that will be used to answer the following research questions:
\begin{itemize}
\item RQ1  What is the effect of each variable on clustering accuracy?
\item RQ2  What is the optimal variable configurations for clustering?
\end{itemize}
\subsection{Experimental Set-up}
\subsubsection{Experimental Subjects.}
The protocol traces included in the experiment are: the \emph{Trivial File Transfer Protocol} (TFTP), the \emph{Domain Name Service} (DNS), \emph{Server Message Block} (SMB), and \emph{Hyper-Text Transfer Protocol} (HTTP). The main datasets have been downloaded from a network security and monitoring website \cite{etresec-2012}. The selected protocols vary in terms of type of data (binary \& text), and the complexity of their message structure. A summary of the collected network traces is provided in table  \ref{var_desc}.

\subsubsection{Experimental Variables.}
The constructed framework enables us to experiment with the following variables: length of the $n$-gram, length of the message, size of the sample, and choice of the distance measure. The four variables have always been key technical questions in the literature \cite{discoverer-2007,Wang3-2011,Wang4-2011,Wang-2012}.

\begin{itemize}
\item \textbf{$N$-gram Length.} We chose a range of values for the $n$-gram for each protocol trace. However, we have also observed the constraint indicated in equations \ref{eq-ng} that the range should not exceed the length of the shortest message in the trace. A summary of the $n$-gram's range for each protocol is shown in table \ref{var_desc} (column 4).

\item \textbf{Message Length.} Three values are selected for the length of the message: \emph{16 bytes}, \emph{32 bytes}, and \emph{64 bytes}. We have experimented with different message lengths ranged from 3 bytes to 64 bytes, we have noticed that clustering scores, for all protocols, tend to be different and erratic when the length of the message is less than 12 bytes, and relatively similar when the length of the message lies between 12 to 16 bytes. We have also noticed that clustering scores gradually decline when the length of the message is greater than 16 bytes.

\begin{table}[t]
\centering
\caption {Summary of network traces and trace-dependant variables.}
\vspace{0.3cm}
\begin{tabular}{l@{\hskip 0.5cm}c@{\hskip 0.5cm}c@{\hskip 0.5cm}|l@{\hskip 0.5cm}l@{\hskip 0.5cm}}
\toprule
Protocol& Sample Size & Type &\multicolumn{2}{c}{Variable} \\
 \cline{4-5} &&& $n$-gram & sub-sample \\
\midrule
	TFTP & 2300 & Binary & 2,3,4 & 500,1000,2000 \\
    DNS  & 4000 & Binary & 2,3,4,5,6,7,8 & 1300,2600,3900 \\ 
    SMB  & 1600 & Mixed  & 2,3,4,5,6,7,8 & 500,1000,1500 \\
    HTTP & 1100 & Text   & 2,3,4,5 & 300,600,900 \\           
\bottomrule
\end{tabular}

\label{var_desc}
\end{table}

\item \textbf{Sample Size.} For each protocol, we have selected three three sub-samples from three different positions of the total sample while maintaining the order of the messages in each sub-sample. The size of each sub-sample is trace dependent and shown in table \ref{var_desc} (column 5). 

\item \textbf{Distance Measure.} With respect to distance measures, we use five distance measures, four  measures are based on the similarity coefficients of the \emph{Jaccard} index, \emph{Dice} index, \emph{Braun-Blanquet} index and the \emph{Cosine} similarity index \cite{Rieck-2008,choi-2010} while the fifth is the \emph{Euclidean} distance measure \cite{Han-2011}. 

For the similarity coefficients, the distance is defined as $D(a,b)=1-S(a,b)$, where $S$ is the similarity of two messages represented by $a$ and $b$ features respectively. The chosen distance measures are diverse and commonly used in the literature \cite{Rieck-2008,comparetti-2009,Wang4-2011}.
\end{itemize}

\subsubsection{Clustering Validation Metrics.} 
We use the \emph{adjusted Rand} index (aka corrected Rand) \cite{Hubert-1985,Meila-2007} as the extrinsic measure. Typically, the score of the adjusted Rand index ranges from 0 to +1 where +1 indicates the two sets of clusters are identical and 0 when the two sets are completely independent. As for intrinsic validation measures, we use the \emph{Ball-Hall} index \cite{ballhall-1965}, \emph{Calinski-Harabasz} index \cite{calinski-1974}, \emph{Davies-Bouldin} index \cite{Davies-1979},  \emph{Trace\_WiB} index \cite{Friedman-1967}, the \emph{SD} index \cite{Halkidi-2001}, and \emph{S\_Dbw} index \cite{sdbw-2001}.

The adjusted Rand index and internal validation indices are chosen based on popularity and recommendations by previous study \cite{int-clust-2010}. We also ruled out internal measures that require intensive calculations.

\subsection{Methodology}
For each protocol trace, we use our framework to cluster protocol messages and validate the results through the extrinsic and intrinsic validation measures. The process is systematically executed using all possible combinations of variable values, and the clustering validation results are recorded each time.
\subsubsection{RQ1. Measuring the Effect of Variables.}
To measure the effect of each variable (e.g., choice of the $n$-gram ), we perform grouped statistical tests on the external validation scores (adjusted Rand). Because we cannot presume normality of the distribution of our data, we resort to \emph{non-parametric} statistical tests. We use \emph{Cohen's} $d$ \footnote{The estimate of $d$ is the statistic denoted by unbiased standardised mean difference or Hedge's g.} \cite{cohen1988,Myers2010} to measure the effect size. 

The basic use of Cohen's $d$ is to measure the mean difference (standardised) between two groups of adjusted Rand scores. Cohen's $d$ is a pairwise test. We carry out every possible pairwise test for each variable (we compare the adjusted Rand scores for every pair of $n$-grams). Because we are mainly interested in the relative distance between variables and not the direction (which one was greater), we take the mean absolute value for all $d$'s to calculate the \emph{aggregate effect} of the variable.

To be able to interpret the magnitude of $d$ for each test, Cohen nominated \emph{0.2}, \emph{0.5} and \emph{0.8} as the \emph{small}, \emph{medium}, and \emph{large} reference values, respectively \cite{cohen1988,cohen1992}. However, Cohen urged researchers to interpret the effect size in the context of their experiments. He offered these reference values only as a "conventional frame of reference" which is recommended when no better basis is available. Typically, the \textit{magnitude} (effect estimate) with the associated \textit{confidence interval} (CI) are reported for each test. We use 95\% as the \textit{confidence level} for all the tests.

\subsubsection{RQ2. Finding Optimal Variable Configurations.}
To answer RQ2, we could simply refer to the highest score returned by the extrinsic measure and retrieve the corresponding variable values. However, in practice, message labels are often not available. Therefore, we use intrinsic validation measures.
Since internal measures can be used to determine the optimal number of clusters \cite{int-clust-2010}, the general procedure to determine the optimal variable configurations is as follows: 
\begin{itemize}
\item Step 1: For each protocol trace, use all possible variable combinations to get different clustering results.
\item Step 2: Measure the clustering result obtained in step 1 using the corresponding internal validation index.
\item Step 3: Choose the best validation result according to the criteria applied with the internal measure. (each internal validation measure has a rule which must be applied in order to obtain the optimal number of clusters). 
\item Step 4: Finally, we retrieve values of variables corresponding to the optimal number of clusters obtained in step 3.
\end{itemize}

\section{Results} 
This sections presents the results of our experiment aiming at illustrating how various variables affect clustering and which configurations lead up to clustering with the highest score using the chosen internal validation measures. 

\subsubsection{RQ1. What is the effect of variables on clustering accuracy?} 
\begin{figure}
\centering
	\subfloat[TFTP]{\includegraphics[scale=0.34]{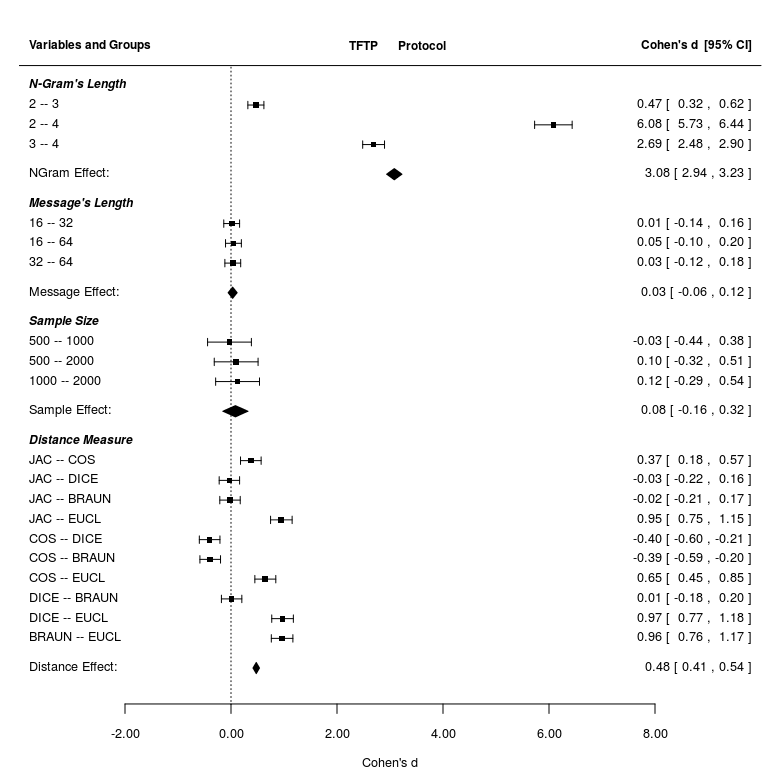}} 
    \subfloat[DNS]{\includegraphics[scale=0.34]{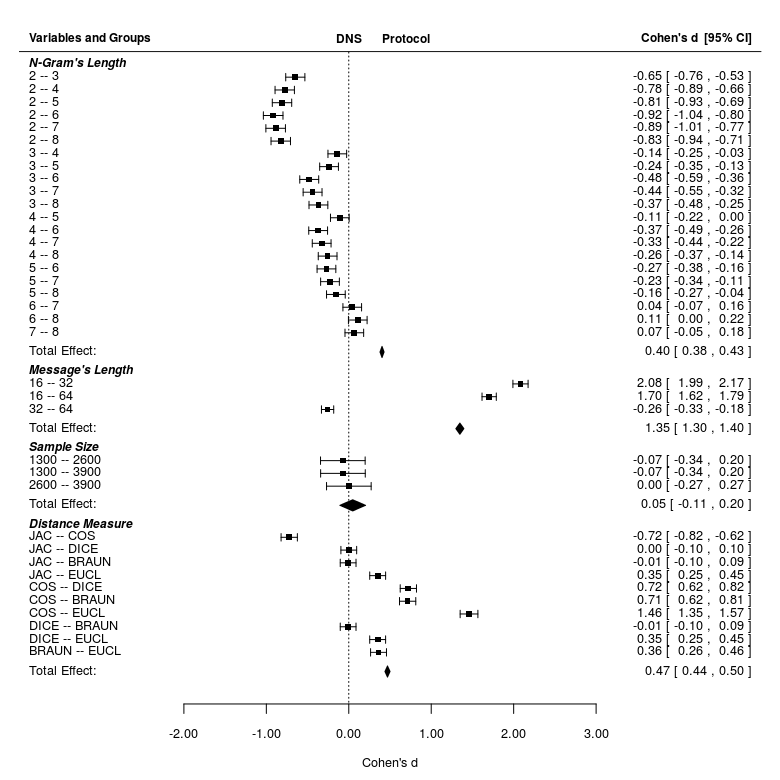}} \\
    \subfloat[SMB]{\includegraphics[scale=0.34]{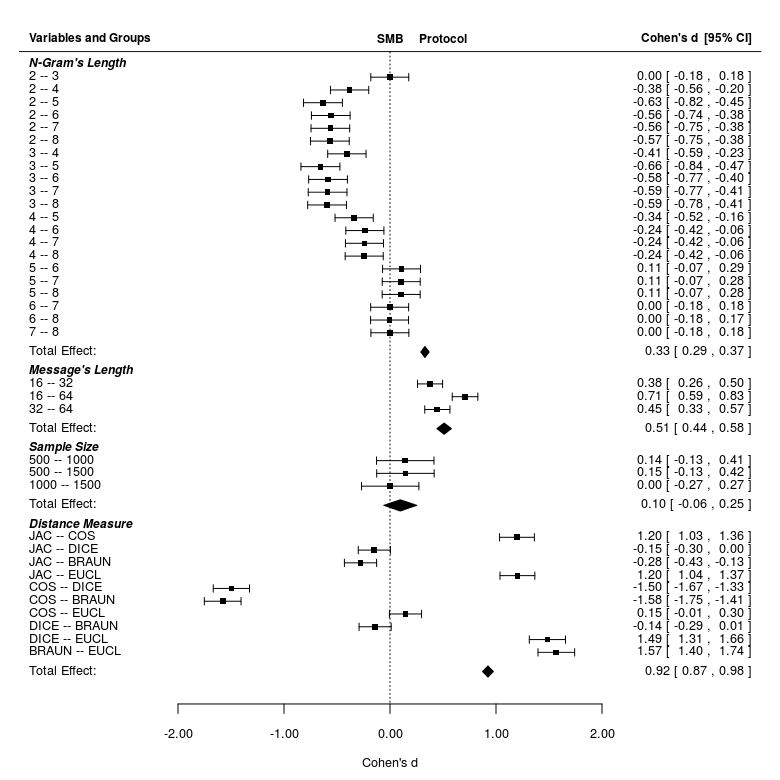}} 
    \subfloat[HTTP]{\includegraphics[scale=0.34]{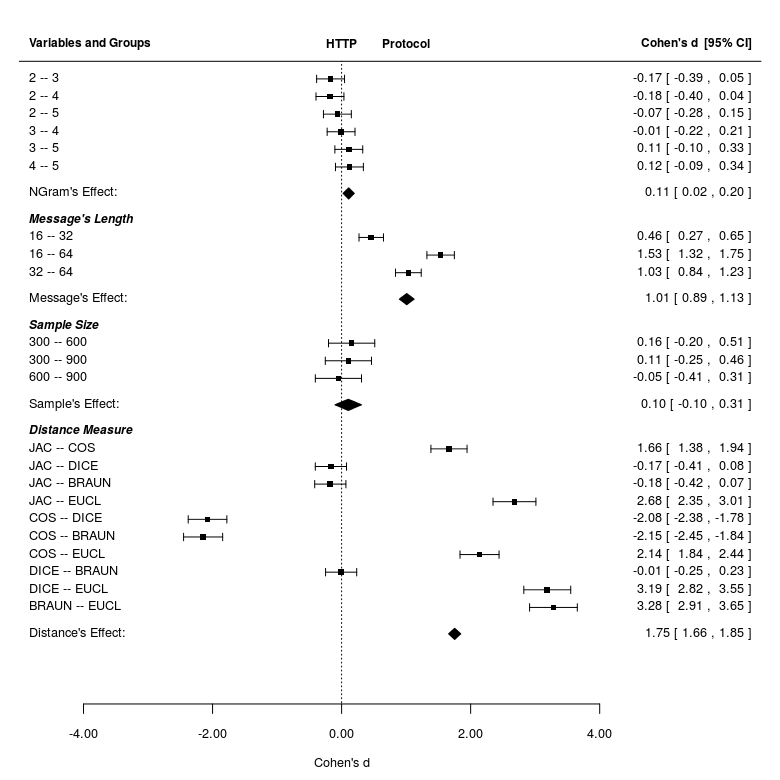}}
\caption{Forest plots showing the effect of variables on clustering accuracy. The figures show the estimated effects of the pairwise tests on the adjusted Rand scores between variable values as well as the aggregate affect of each variable. It also, shows the corresponding 95\% confidence intervals for each test.}
\label{Forset_Clust}
\end{figure}

Figure \ref{Forset_Clust} (a-d) shows Forest plots illustrating the effect of variables. The left-hand column lists the names of the variables and pairwise tests carried out between variable values. The right-column is a plot of these effects (shown as squares) within confidence intervals represented as horizontal lines. The overall effect of each variable is shown as a diamond. A vertical line indicating no-effect is also plotted. 

The overall results show that the distance measure and the length of the message have a significant effect on clustering accuracy. Therefore, the choices of these variables are important. However, for TFTP, the choice of the $n$-gram seems to be the pivotal variable for clustering. The overall effect of the sample size is negligible. The results are explained in more detail below.

\begin{itemize}
\item \textbf{Distance Measure.} The effect of the distance measure on SMB \& HTTP is significantly \emph{large}. The overall effect is clearly visible in Figure \ref{Forset_Clust} (c-d) as point estimates confidence intervals are shifted away from the no-effect line. The effect of the distance measure on TFTP \& DNS is relativity less (\emph{medium}). However, judging the precision of the estimated effects corroborated by the short confidence intervals, the effect is big enough to indicate the importance of the distance measure for all protocols.

\item \textbf{Message Length.} For DNS \& HTTP, the average effect of the message's length is greater than one std. (standard deviation) which is very \emph{large} for both protocols, while the effect on SMB is about 0.5 std. (\emph{medium}). For the TFTP protocol, the message length does not seem to have any effect on clustering, this is clearly indicated in Figure \ref{Forset_Clust} (a) as all effect estimates lie on the no-effect line.
\item \textbf{$N$-gram Length.} The effect of the $n$-gram on DNS, SMB and HTTP ranges from \emph{small} to \emph{medium}. However, Figure \ref{Forset_Clust} (b\&c) indicate that the effect of this variable is much more significant for the DNS \& SMB protocols than the HTTP as clearly shown by the individual tests as well as the overall effect of the variable. As for TFTP, the effect of the $n$-gram is critically (\emph{large}) with 3.08 standard deviation. Therefore, length of the $n$-gram is important choice.

\item \textbf{Sample Size.} The overall effect of the sample size is \emph{negligible}. For all protocols, this is clearly evident that the effect of this variable lies within wider confidence intervals and all of these confidence intervals intersect with the no-effect line which indicates that sample size as a whole has insignificant impact on clustering.
\end{itemize}
\subsubsection{RQ2: What is the optimal variable configurations for clustering?} 

In general, the results show that the combination of the \emph{Ball-Hall} validity index and the \emph{Braun-Blanquet} binary similarity measure tend to give the best results in predicting the optimal variable configuration for clustering.

The results are shown in  Table \ref{tbl_best_combo_int} (a-d). The table shows the experimental variables and chosen internal measures as well the the score of the adjusted Rand corresponding to each internal measure. For TFTP and DNS, the \emph{Ball-Hall} index has predicted the best variable combination (clustering score) as indicated in Table \ref{tbl_best_combo_int} (a-b), while the \emph{SD} and \emph{Calinski-Harabasz} indices have predicted the best variable combination for SMB and HTTP protocols respectively with the \emph{Ball-Hall} index comes the second. 

Also, Table \ref{tbl_best_combo_int} (a-d) shows internal measures tend to  give better predictions with the binary similarity measures of Braun-Blanquet, Dice \& Jaccard.

\begin{table}
  \centering
  \caption {Performance of internal validation measures in predicting optimal variable configuration for clustering.}
  \vspace{0.5cm}
  \scalebox{0.58}{\subfloat[TFTP]{\definecolor{Gray}{gray}{0.85}
\begin{tabular}{cccccl}
\toprule
Distance & Sample & $n$-gram & Message & adj. Rand & Internal Measure \\
\midrule
\multirow {6}{*}{Jaccard}    & 200  & 2 & 16 & 0.998782 & Trace\_WiB \\ 
  							  & 2300 & 2 & 16 & 0.998254 & Ball\_Hall \\ 
							  & 100  & 3 & 16 & 0.938577 & SD\_Dis \\ 
							  & 1500 & 4 & 16 & 0.000214 & Calinski\_Harabasz \\ 
							  & 1500 & 4 & 16 & 0.000214 & Davies\_Bouldin \\ 
							  & 1100 & 4 & 16 & 0.000131 & S\_Dbw \\ 
\hline					              
\multirow {6}{*}{Dice} 	& 200  & 2 & 16 & 0.998782 & Trace\_WiB \\ 
					    & 2300 & 2 & 32 & 0.998254 & Ball\_Hall \\ 
  						& 100  & 3 & 16 & 0.938577 & SD\_Dis \\ 
					 	& 1500 & 4 & 16 & 0.000214 & Calinski\_Harabasz \\ 
					    & 1500 & 4 & 16 & 0.000214 & Davies\_Bouldin \\ 
					    & 1100 & 4 & 16 & 0.000131 & S\_Dbw \\  
\hline		
\rowcolor{Gray}			 
\multirow {6}{*}{Braun-Blanquet}  & 2300 & 2 & 32 & 0.999982 & Ball\_Hall \\ 
\rowcolor{Gray}	
								  & 2300 & 2 & 32 & 0.999982 & Trace\_WiB \\
  								  & 100  & 3 & 16 & 0.938577 & SD\_Dis \\ 
  								  & 1500 & 4 & 16 & 0.000214 & Calinski\_Harabasz \\ 
								  & 1500 & 4 & 16 & 0.000214 & Davies\_Bouldin \\ 
								  & 1100 & 4 & 16 & 0.000117 & S\_Dbw \\ 
\hline
					        
\multirow {6}{*}{Cosine} 	& 1100     & 3  & 16 & 0.945071 &S\_Dbw  \\
	    				   	& 800      & 2  & 64 & 0.629911 & Trace\_WiB \\
						    & 100      & 2  & 16 & 0.104278 & SD\_Dis \\ 
					        & 1700     & 4  & 64 & 0.003506 & Ball\_Hall \\ 
					        & 1500     & 4  & 16 & 0.000214 & Calinski\_Harabasz \\   
					        & 1500     & 4  & 16 & 0.000214 & Davies\_Bouldin \\ 
\hline 					       
\multirow {6}{*}{Euclidean}  & 100  & 4 & 32 & 0.314674 & Ball\_Hall \\
						      & 2100 & 3 & 32 & 0.015229 & S\_Dbw \\  
   						      & 1200 & 4 & 32 & 0.000971 & Trace\_WiB \\
   						      & 800  & 4 & 16 & 0.000610 & SD\_Dis \\  
   						      & 2200 & 4 & 16 & 0.000230 & Calinski\_Harabasz \\ 
                             & 1500 & 4 & 16 & 0.000000 & Davies\_Bouldin \\

\bottomrule       
\end{tabular}
}} 
  \scalebox{0.58}{\subfloat[DNS]{
\definecolor{Gray}{gray}{0.85}
\begin{tabular}{cccccl}
\toprule
Distance & Sample & $n$-gram & Message & adj. Rand & Internal Measure \\
\midrule
\rowcolor{Gray}	
\multirow {6}{*}{Jaccard} 	&4000 & 7 & 16 & 0.989203 & Ball\_Hall \\ 
 							&4000 & 4 & 16 & 0.531224 & Calinski\_Harabasz \\ 
							&200 & 4 & 16 & 0.344495 & SD\_Dis \\ 
  							&2300 & 2 & 16 & 0.111911 & Davies\_Bouldin \\ 
						    &100 & 2 & 16 & 0.072107 & S\_Dbw \\ 
						    &800 & 2 & 32 & 0.019798 & Trace\_WiB \\ 
\hline
\rowcolor{Gray}						              
\multirow {6}{*}{Dice}	 	&4000 & 7 & 16 & 0.989203 & Ball\_Hall \\ 
  							&4000 & 4 & 16 & 0.531224 & Calinski\_Harabasz \\ 
						    &200  & 4 & 16 & 0.344495 & SD\_Dis \\ 
  						    &2300 & 2 & 16 & 0.111911 & Davies\_Bouldin \\ 
						    &100  & 2 & 16 & 0.072107 & S\_Dbw \\ 
						    &800  & 2 & 32 & 0.019798 & Trace\_WiB \\ 
\hline
\rowcolor{Gray}					        
\multirow {6}{*}{Braun-Blanquet} & 4000 & 7 & 16 & 0.989203 & Ball\_Hall \\ 
 								 &4000 & 4 & 16 & 0.531224 & Calinski\_Harabasz \\ 
							     &200 & 4 & 16 & 0.344495 & SD\_Dis \\ 
							     &2300 & 2 & 16 & 0.189199 & Davies\_Bouldin \\ 
							     &400 & 2 & 16 & 0.077443 & Trace\_WiB \\ 
							     &100 & 2 & 16 & 0.072107 & S\_Dbw \\ 
\hline				        
\multirow {6}{*}{Cosine} 	&4000 & 6 & 16 & 0.945626 & Ball\_Hall \\ 
  							&3600 & 3 & 32 & 0.346439 & Trace\_WiB \\ 
						    &300 & 8 & 16 & 0.297692 & SD\_Dis \\ 
						    &400 & 2 & 16 & 0.134757 & S\_Dbw \\ 
						    &4000 & 2 & 16 & 0.116275 & Davies\_Bouldin \\ 
						    &4000 & 2 & 32 & 0.044579 & Calinski\_Harabasz \\ 
\hline				       
\multirow {6}{*}{Euclidean}  &2300 & 8 & 16 & 0.233399 & Calinski\_Harabasz \\ 
  							  &1900 & 8 & 16 & 0.211938 & SD\_Dis \\ 
							  &800 & 7 & 16 & 0.175221 & Trace\_WiB \\  
							  &300 & 3 & 64 &  0.027146 & Ball\_Hall \\ 
							  &3400 & 3 & 16 & 0.012933 & S\_Dbw \\ 
   							  &100 & 2 & 32 & 0.000000 & Davies\_Bouldin \\ 
\bottomrule 	       
\end{tabular}
}} \\
  \scalebox{0.58}{\subfloat[SMB]{\definecolor{Gray}{gray}{0.85}
\begin{tabular}{cccccl}
\toprule
Distance & Sample & $n$-gram & Message & adj. Rand & Internal Measure \\
\midrule
\rowcolor{Gray}	
\multirow {6}{*}{Jaccard} 	& 100 & 5 & 16 & 1.000000 & SD\_Dis \\ 
 						    &100 & 4 & 64 & 0.992056 & Trace\_WiB \\ 
						    &1600 & 3 & 16 & 0.180173 & Ball\_Hall \\ 
							&1600 & 5 & 32 & 0.002438 & Calinski\_Harabasz \\ 
							&1600 & 3 & 32 & 0.002438 & S\_Dbw \\ 
						    &100 & 2 & 64 & 0.000000 & Davies\_Bouldin \\ 
\hline
\rowcolor{Gray}						              
\multirow {6}{*}{Dice}  & 100 & 5 & 16 & 1.000000 & SD\_Dis \\ 
						    &100 & 4 & 64 & 0.710890 & Trace\_WiB \\ 
						    &1600 & 3 & 16 & 0.180173 & Ball\_Hall \\ 
							&1600 & 5 & 32 & 0.002438 & Calinski\_Harabasz \\ 
							&1600 & 3 & 32 & 0.002438 & S\_Dbw \\ 
							&100 & 2 & 64 & 0.000000 & Davies\_Bouldin \\ 
\hline
\rowcolor{Gray}	
\multirow {6}{*}{Braun-Blanquet} & 100 & 5 & 16 & 1.000000 & SD\_Dis \\ 
										  &1600 & 4 & 32 & 0.747453 & Ball\_Hall \\ 
										  &100 & 4 & 64 & 0.710890 & Trace\_WiB \\ 
										  &1600 & 2 & 16 & 0.002438 & Calinski\_Harabasz \\ 
										  &1600 & 2 & 16 & 0.002438 & S\_Dbw \\ 
										  &100 & 2 & 64 & 0.000000 & Davies\_Bouldin \\ 
\hline					        
\multirow {6}{*}{Cosine} 	&100 & 5 & 16 & 0.271637 & SD\_Dis \\ 
  							&200 & 8 & 16 & 0.158804 & Calinski\_Harabasz \\ 
						    &200 & 8 & 16 & 0.158804 & Davies\_Bouldin \\ 
						    &200 & 8 & 16 & 0.158804 & S\_Dbw \\ 
						    &300 & 5 & 16 & 0.047732 & Ball\_Hall \\ 
						    &1000 & 4 &16 & 0.003836 & Trace\_WiB \\ 
\hline					       
\multirow {6}{*}{Euclidean} &1600 & 7 & 64 & 0.000000 & Ball\_Hall \\ 
  							&400 & 2 & 16 & 0.000000 & Calinski\_Harabasz \\ 
  							&100 & 2 & 16 & 0.000000 & Davies\_Bouldin \\ 
						    &700 & 2 & 16 & 0.000000 & SD\_Dis \\ 
						    &1200 & 7 & 16 & 0.000000 & S\_Dbw \\ 
						    &200 & 2 & 16 & 0.000000 & Trace\_WiB \\ 
\bottomrule        
\end{tabular}
}} 
  \scalebox{0.58}{\subfloat[HTTP]{\definecolor{Gray}{gray}{0.85}
\begin{tabular}{cccccl}
\toprule 
Distance & Sample & $n$-gram & Message & adj. Rand & Internal Measure \\
\midrule 
\multirow {6}{*}{Jaccard}    &1000 & 5 & 16 & 0.941975 & Calinski\_Harabasz \\ 
 							  &900 & 5 & 16 & 0.941237 & S\_Dbw \\ 
							  &1100 & 4 & 16 & 0.924386 & Ball\_Hall \\ 
							  &100 & 5 & 16 & 0.821678 & SD\_Dis \\ 
							  &1100 & 5 & 32 & 0.448247 & Trace\_WiB \\ 
							  &700 & 3 & 64 & 0.335951 & Davies\_Bouldin \\ 
\hline					              
\multirow {6}{*}{Dice} 		&1000 & 5 & 16 & 0.941975 & Calinski\_Harabasz \\ 
							&900 & 5 & 16 & 0.941237 & S\_Dbw \\ 
							&1100 & 4 & 16 & 0.924386 & Ball\_Hall \\ 
							&100 & 5 & 16 & 0.821678 & SD\_Dis \\ 
							&700 & 3 & 64 & 0.471805 & Davies\_Bouldin \\ 
							&100 & 2 & 64 & 0.238019 & Trace\_WiB \\ 
\hline
\rowcolor{Gray}						 
\multirow {6}{*}{Braun-Blanquet}  & 1000 & 5 & 16 & 0.942082 & Calinski\_Harabasz \\ 
										    &1100 & 4 & 16 & 0.902149 & Ball\_Hall \\ 
											&100 & 5 & 16 & 0.821678 & SD\_Dis \\ 
											&100 & 2 & 16 & 0.764644 & S\_Dbw \\ 
											&700 & 3 & 64 & 0.453987 & Davies\_Bouldin \\ 
											&100 & 4 & 64 & 0.403513 & Trace\_WiB \\
\hline
\multirow {6}{*}{Cosine} 		&800 & 4 & 16 & 0.465563 & Calinski\_Harabasz \\ 
								&100 & 3 & 16 & 0.431648 & Ball\_Hall \\ 
								&100 & 3 & 16 & 0.431648 & Davies\_Bouldin \\ 
								&200 & 3 & 16 & 0.362845 & S\_Dbw \\ 
								&100 & 5 & 16 & 0.355834 & SD\_Dis \\ 
								&100 & 2 & 32 & 0.192357 & Trace\_WiB \\ 
\hline				       
\multirow {6}{*}{Euclidean}  &1000 & 5 & 16 & 0.311167 & Calinski\_Harabasz \\ 
							  &700 & 3 & 16 & 0.274333 & S\_Dbw \\ 
							  &100 & 3 & 16 & 0.145950 & Ball\_Hall \\ 
							  &100 & 5 & 32 & 0.030939 & SD\_Dis \\
							  &100 & 2 & 64 & 0.000000 & Davies\_Bouldin \\ 
							  &100 & 4 & 64 & 0.000000 & Trace\_WiB \\  
\bottomrule 
\end{tabular}
}}
    \label{tbl_best_combo_int}
\end{table}

\section{Threats to Validity} 
Although all experiments were tested on exactly the same machine and under the same experimental configurations, threats to \emph{external validity} might arise which might limit the generalisability of these findings.

\begin{itemize}
\item \emph{Representative Protocols.} Since we our study involved only four network protocols, they may not be representative of the entire family of network protocols . However, this threat is partially considered by selecting the possible types of network protocols (text \& binary protocols).

\item \emph{Representative Traces.} Some of the collected network traces are relatively small in size and may not be representative of the protocol under study. The effect of some of the variables for the TFTP protocol vary from the rest of the protocols (DNS,SMB \& HTTP), this is could be due to the fact that the gathered messages are not well trained to be representative of the protocol behaviour (lack of diversity of traffic seen in the trace).
\end{itemize}

\section{Conclusions and Future Work}

In this paper, we investigated the impact of four important variables on clustering accuracy as part of reverse engineering protocols from network traces. To support our investigation, we have developed a modular framework that enables us to produce arbitrary clustering configurations of protocol inferencing. We have applied this framework to data traces from four widely used network protocols. Our research indicates the following:
\begin{itemize}
\item The choice of the distance measure and length of the message is of paramount importance for clustering.
\item The number of messages in the trace does not have significant impact on clustering accuracy.

\item It is possible to derive highly accurate clustering configurations without relying upon labelled examples (i.e., by using internal validation measures). 
\end{itemize}

In the future, we plan to mitigate threats to validity by incorporating more diverse network protocols. We also plan to enrich our protocol inference by integrating clustering internal validation to predict optimal configurations for clustering.

\bibliographystyle{splncs03}
\bibliography{ref}


\end{document}